\documentclass[12pt,preprint]{aastex}
\usepackage{psfig}

\def\apj{ApJ} 
\def\apjl{ApJL} 
\def\mnras{MNRAS} 
\def\pasp{PASP}

\def\araa{ARAA} 
\def\aap{A\&AP} 
 
\def\apjs{APJS} 
 
\def\nat{Nature}
\def\PhysRep{Physics Reports}
\def\aa{A\&A}
\def\science{Science}

\def\gs{\mathrel{\raise0.35ex\hbox{$\scriptstyle >$}\kern-0.6em\lower0.40ex\hbox{{$\scriptstyle \sim$}}}} 
\def\ls{\mathrel{\raise0.35ex\hbox{$\scriptstyle <$}\kern-0.6em\lower0.40ex\hbox{{$\scriptstyle \sim$}}}}

\def\Wm2{\,\hbox{W}\,\hbox{m}^{-2}} 
\def\gsim{\mathrel{\raise0.35ex\hbox{$\scriptstyle >$}\kern-0.6em\lower0.40ex\hbox{{$\scriptstyle \sim$}}}} 
\def\lsim{\mathrel{\raise0.35ex\hbox{$\scriptstyle <$}\kern-0.6em\lower0.40ex\hbox{{$\scriptstyle \sim$}}}} 
\def\ltsima{$\; \buildrel < \over \sim \;$} 
\def\simlt{\lower.5ex\hbox{\ltsima}} 
\def\gtsima{$\; \buildrel > \over \sim \;$} 
\def\simgt{\lower.5ex\hbox{\gtsima}}

\shorttitle{Gas and dust in a $z=4.24$ submillimeter galaxy}
\shortauthors{Cox et al.}

\begin{document}

\title{Gas and dust in a submillimeter galaxy at {\it z} = 4.24 from the {\it Herschel} ATLAS}

\author{P.\ Cox\altaffilmark{1}, 
M.\ Krips\altaffilmark{1}, 
R.\ Neri\altaffilmark{1},
A.\ Omont\altaffilmark{2},
R.\ G\"usten\altaffilmark{3}, 
K.\,M.\ Menten\altaffilmark{3}, 
F.\ Wyrowski\altaffilmark{3}, 
A.\ Wei\ss\altaffilmark{3},
A.\ Beelen\altaffilmark{4},
M.\,A.\ Gurwell\altaffilmark{5},
H.\ Dannerbauer\altaffilmark{6},
R.\,J.\ Ivison\altaffilmark{7,8},
M.\ Negrello\altaffilmark{9},
I.\ Aretxaga\altaffilmark{10},
D.\,H.\ Hughes\altaffilmark{10},
R.\ Auld\altaffilmark{11},
M.\ Baes\altaffilmark{12},
R.\ Blundell\altaffilmark{5},
S.\ Buttiglione\altaffilmark{13},
A.\ Cava\altaffilmark{14},
A.\ Cooray\altaffilmark{15},
A.\ Dariush\altaffilmark{11,16},
L.\ Dunne\altaffilmark{17},
S.\ Dye\altaffilmark{17},
S.\,A.\ Eales\altaffilmark{11},
D.\ Frayer\altaffilmark{18},
J.\ Fritz\altaffilmark{12},
R.\ Gavazzi\altaffilmark{2},
R.\ Hopwood\altaffilmark{9},
E.\ Ibar\altaffilmark{7},
M.\ Jarvis\altaffilmark{13},
S.\ Maddox\altaffilmark{17},
M.\ Micha{\l}owski\altaffilmark{8},
E.\ Pascale\altaffilmark{11},
M.\ Pohlen\altaffilmark{11},
E.\ Rigby\altaffilmark{17},
D.\,J.\,B.\ Smith\altaffilmark{17},
A.\,M.\ Swinbank\altaffilmark{19},
P.\ Temi\altaffilmark{20},
I.\ Valtchanov\altaffilmark{21},
P.\ van der Werf\altaffilmark{22}
and G.\ de Zotti\altaffilmark{13}
} 

\altaffiltext{1}{IRAM, 300 rue de la piscine, F-38406 Saint-Martin d'H\`eres, France; e-mail: cox@iram.fr}
\altaffiltext{2}{Institut d'Astrophysique de Paris, Universit\'e Pierre et Marie Curie and CNRS (UMR7095), 98 bis boulevard
Arago, 75014 Paris, France}
\altaffiltext{3}{MPIfR, Auf dem H\"ugel 69, 53121 Bonn, Germany}
\altaffiltext{4}{IAS, B\^at.\ 121, Universit\'e Paris-Sud, F-91405 Orsay Cedex, France}
\altaffiltext{5}{Harvard-Smithsonian Center for Astrophysics, 60 Garden Street, Mail Stop 42, Cambridge, MA 02138, USA}
\altaffiltext{6}{AIM, CEA/DSM-CNRS-Universit\'e de Paris Diderot, DAPNIA/Service d'Astrophysique, CEA Saclay, Orme des Merisiers, 91191 Gif-sur-Yvette, France}
\altaffiltext{7}{UK Astronomy Technology Centre, Royal Observatory, Blackford Hill, Edinburgh EH9 3HJ, UK}
\altaffiltext{8}{Scottish Universities Physics Alliance, Institute for Astronomy, University of Edinburgh, Royal Observatory, Blackford Hill, Edinburgh EH9 3HJ, UK}
\altaffiltext{9}{Department of Physics and Astronomy, The Open University, Milton Keynes MK7 6AA, UK}
\altaffiltext{10}{Instituto Nacional de Astrof\'isica, Optica y Electr\'onica, Aptdo.\ Postal 51 y 216, 72000 Puebla, Mexico}
\altaffiltext{11}{School of Physics and Astronomy, Cardiff University, The Parade, Cardiff CF24 3AA, UK}
\altaffiltext{12}{Sterrenkundig Observatorium, Universiteit Gent, Krijgslaan 281, S9, B-9000 Gent, Belgium}
\altaffiltext{13}{INAF-Osservatorio Astronomico di Padova, Vicolo
  dell'Osservatorio 5, I-35122 Padova, Italy \& SISSA, Via Bonomea
  265, I-34136 Trieste, Italy}
\altaffiltext{14}{Departamento de Astrof\'{\i}sica, Facultad de CC. F\'{\i}sicas,
Universidad Complutense de Madrid, E-28040 Madrid, Spain}
\altaffiltext{15}{Department of Physics and Astronomy, University of California, Irvine, CA 92697, USA}
\altaffiltext{16}{School of Astronomy, Institute for Research in Fundamental Sciences
(IPM), P.O.\ Box 19395-5746, Tehran, Iran}
\altaffiltext{17}{School of Physics and Astronomy, University of Nottingham, University Park, Nottingham NG7 2RD, UK}
\altaffiltext{18}{National Radio Astronomy Observatory, P.O.\ Box 2, Green Bank, WV 24944, USA}
\altaffiltext{19}{Institute for Computational Cosmology, Durham University, South Road, Durham DH1 3EE, UK}
\altaffiltext{20}{Astrophysics Branch, NASA Ames Research Center, Mail Stop 245-6, Moffett Field, CA 94035, USA}
\altaffiltext{21}{Herschel Science Center, ESAC, ESA, P.O.\ Box 78, 28691 Villanueva de la Ca\~nada, Madrid, Spain}
\altaffiltext{22}{Leiden Observatory, Leiden University, P.O.\ Box 9513, 2300 RA Leiden, The Netherlands}

\begin{abstract}
We report ground-based follow-up observations of the exceptional
source, ID\,141, one the brightest sources detected so far in the {\it
  H}-ATLAS cosmological survey. ID\,141 was observed using the IRAM
30-meter telescope and Plateau de Bure interferometer (PdBI), the
Submillimeter Array (SMA) and the Atacama Pathfinder Experiment (APEX)
submillimeter telescope to measure the dust continuum and emission
lines of the main isotope of carbon monoxide and carbon ([C\,{\small
    I}] and [C\,{\small II}]). The detection of strong CO emission
lines with the PdBI confirms that ID\,141 is at high redshift ($z=4.243
\pm 0.001$).  The strength of the continuum and emission lines
suggests that ID\,141 is gravitationally lensed.  The width ($\Delta
V_{\rm FWHM} \sim 800 \, \rm km \, s^{-1}$) and asymmetric profiles of
the CO and carbon lines indicate orbital motion in a disc or a
merger. The properties derived for ID\,141 are compatible with a
ultraluminous ($L_{\rm FIR} \sim 8.5\pm0.3 \times 10^{13} \, \mu_{\rm
  L}^{-1} \, \rm L_\odot$, where $\mu_L$ is the amplification factor),
dense ($n \approx 10^4 \, \rm cm^{-3}$) and warm ($T_{\rm kin} \approx 40
\, \rm K$) starburst galaxy, with an estimated star-formation rate of
(0.7 to 1.7)$\times 10^4 \, \mu_{\rm L}^{-1} \rm \, M_\odot / yr$. The
carbon emission lines indicate a dense ($n \approx 10^4 \rm cm^{-3}$)
Photo-Dominated Region, illuminated by a far-UV radiation field a few
thousand times more intense than that in our Galaxy. In conclusion,
the physical properties of the high-$z$ galaxy ID\,141 are remarkably
similar to those of local ultraluminous infrared galaxies.
\end{abstract}


\section{Introduction}

Over the last decade, submillimeter surveys have revolutionized our
understanding of the formation and evolution of galaxies by uncovering
a population of high-redshift, dust-obscured systems that are forming
stars at a tremendous rate (e.g.\ Smail et al.\ 1997; Hughes et
al.\ 1998; Blain et al.\ 2002).  Submillimeter galaxies (SMG) are
high-redshift analogues of the local ultraluminous infrared galaxies
(ULIRGs) in terms of their CO, radio and infrared properties (Tacconi
et al.\ 2008, 2010). The general importance of ultraluminous sources
for galaxy evolution has been highlighted by recent {\it Spitzer} and
{\it Herschel} results that show that these sources contribute
significantly to the total amount of star formation in the early
universe \cite{Magnelli2009}.

Until recently, observations at submillimeter wavelengths have been
conducted mainly from the ground and/or from balloon experiments and
covered only relatively small areas of the sky. The discovery rate of
rare objects such as strongly lensed sources was therefore
limited. Nevertheless, due to selection effects, many of the known
high-redshift submillimeter sources are gravitationally lensed, such
as {\it IRAS}\,F10214+4724, which is amplified by a factor of $\sim
10\times$ \cite{Rowan-Robinson1991, Downes1995}.  However, these
limited surveys found few strongly lensed sources.  This situation
changed dramatically with the 87\,deg$^2$ deep field survey at
wavelengths of 1.4 and 2.0~mm taken with the South Pole Telescope
\cite{Vieira2010} and, since then, the {\it Herschel} space
observatory \cite{Pilbratt2010} has been able to map an enormously
extended area of the sky at submillimeter wavelengths.  In particular,
the wide-area {\it Herschel} Astrophysical Terahertz Large Area Survey
or {\it H}-ATLAS \cite{Eales2010} with the SPIRE \cite{Griffin2010}
and the PACS \cite{Poglitsch2010} instruments will ultimately map a
total of about 570 square-degrees to around (or below) the confusion
limit in five bands from 100 to 500~$\rm \mu m$. In addition to
uncovering hundreds of thousands of SMGs, the {\it Herschel} surveys
will enable discovery of the brightest infrared sources in the
universe, including many strongly lensed sources because of the high
magnification bias in the submillimeter bands \cite{Blain1996}. These
surveys and samples of lensed sources are unique and will remain so
for the foreseeable future.

Negrello et al.\ (2007) predicted that most of the brightest 500-$\rm
\mu m$ sources (i.e.\ those with $S_{\rm 500 \, \mu m} > 100$\,mJy)
detected at high redshift by the {\it Herschel} space observatory will
consist of strongly lensed submillimeter galaxies, a population of
local star-forming galaxies -- easily identified in shallow Sloan
Digital Sky Survey (SDSS) images -- a relatively small fraction of
radio-bright AGNs -- easily identifiable in shallow radio surveys
\cite{Negrello2010} -- as well as nearby spirals.  The ongoing surveys
have confirmed this prediction with the detection of
submillimeter-bright ($S_{\rm 500 \, \mu m} > 100$\,mJy) sources that
have been shown, based on ground-based follow-up observations, to be
gravitationally lensed SMGs in the distant ($z \sim 1.5-4$) Universe
\cite{Negrello2010, Conley2011}.  The lensing boosts the sensitivity
of observations and improves their spatial resolution enabling
detailed studies of the populations responsible for the bulk of the
background \cite{Blain1999, Hopwood2010}.  Follow-up observations with
ground-based telescopes, especially by (sub)millimeter facilities are
essential to further probe the physical characteristics of these
outstanding sources.  A recent and striking example of a lensed source
discovered in the pre-{\it Herschel} era in the submillimeter is the
$z=2.3$ galaxy SMM\,J2135$-$0102 \cite{Swinbank2010} or the `Cosmic
Eyelash'. This typical SMG, which is amplified by a factor of
$32.5\times$, has been observed in exquisite detail enabling diagnosis
of the properties of its molecular and atomic gas \cite{Danielson2011,
  Swinbank2011}.

Here, we report ground-based follow-up observations of an exceptional
source uncovered by the {\it H}-ATLAS survey,
HATLAS\,J142413.9+022304 (hereafter ID\,141), including a measurement
of its redshift and a derivation of the properties of its molecular
and atomic gas content.
 
Throughout the paper we use a $\Lambda$CDM cosmology with $H_0 =
71$\,km\,s$^{-1}$\,Mpc$^{-1}$, $\Omega_m=0.27$ and
$\Omega_{\Lambda}=1-\Omega_m$ \cite{Spergel03}.

\section{Observations and results}

The source ID\,141 is one of the brightest sources detected so far in
the {\it Herschel} deep surveys. It is the strongest source peaking at
500~$\rm \mu m$ in the 100~deg$^2$ currently surveyed in the {\it
  H}-ATLAS program and yet found in the {\it Herschel} cosmological
surveys. The source position as determined by {\it Herschel} is RA
14:24:13.9, Dec.\ +02:23:04 (J2000) and is located in the so-called
`GAMA 15h' field. Its flux densities, as measured with SPIRE, are in
excess of 100~mJy in all bands with a spectral energy distribution
(SED) that is rising from 250 to 500~$\rm \mu m$
(Table~\ref{table:photometry}), indicating that the redshift of this
exceptional source must be high ($z>3$).

Imaging from the SDSS (DR7 - York et al.\ 2000) reveals a faint
optical counterpart with the following magnitudes $u=23.45\pm 1.33$,
$g=22.99\pm 0.35$, $r=22.06\pm 0.23$, $i=20.79\pm 0.12$, and
$z=20.35\pm 0.40$, that is located within 0.4$^{\prime\prime}$ of the
far-infrared position as measured by {\it Herschel}. The photometric
redshift based on the optical magnitudes is $z=0.69\pm 0.13$, which is
clearly incompatible with the high redshift indicated by the
far-infared photometry and more consistent with the hypothesis that
this source is the lensing galaxy \cite{Negrello2010}.

\subsection{30-meter Telescope}

ID\,141 was subsequently observed at the IRAM 30-meter telescope on
2010 May 19 using the 117-channel bolometer array MAMBO-2 to measure
the dust continuum flux density at 1.2~mm. The observations were made
using the on--off mode, wobbling by $32\arcsec$ in azimuth at a rate
of 2~Hz. The target source was positioned on the most sensitive
bolometer, and the correlated sky noise was determined from the other
bolometers and subtracted from the `on-source bolometer'. The source
was observed twice, each time for 10~min, yielding a final
r.m.s.\ noise level of $\rm \sim 2 \, mJy$.  Standard sources (planets
and secondary calibration sources) were used for absolute flux
calibration.  The data were reduced using the standard scripts for
on--off observation data in the MOPSIC software package
\cite{Zylka1988}. The measured flux density of ID\,141 is $S_{\rm
  1.2mm} = 36 \pm 2 mJy$, making it one of the high-redshift sources
with the largest flux density measured to date at millimeter
wavelengths.

Together with the {\it Herschel} data, the 1.2-mm measurement
indicates that the photometric redshift is in the range $3.3 < z_{\rm
  phot} < 4.5$ -- following the method of fitting low-redshift
templates to the far-infrared/millimeter available data described in
Aretxaga et al.\ (2005) -- a range containing 68\% of the probability
distribution that corresponds to a very flat distribution due to the
lack of photometric data between $\rm 500 \, \mu m$ and 1.2~mm, where
the peak of the spectral energy distribution (SED) is lying.

\subsection{Plateau de Bure Interferometer (PdBI)}

Based on the redshift range estimated from the available photometric
data of {\it Herschel} and MAMBO, we used the IRAM PdBI (with five
antennas) to search for CO emission lines and determine the redshift
of ID\,141.  This search was made possible by the recently installed
wide-band correlator WideX that provides a contiguous frequency
coverage of 3.6~GHz bandwidth in dual polarization with a fixed
channel spacing of 1.95~MHz. The observations were made in D
configuration from 2010 May 23 to May 27 in a series of brief
integrations with good atmospheric phase stability (seeing of $1''$)
and reasonable transparency (PWV of 5--10~mm). Absolute fluxes was
calibrated on MWC~349 when available or the quasar 1502+1749 whose
flux is regularly monitored at the IRAM PdBI. The accuracy of the flux
calibration is estimated to be $\sim$10\% at 3 and 2~mm.

\begin{figure*} 
\centerline{\psfig{figure=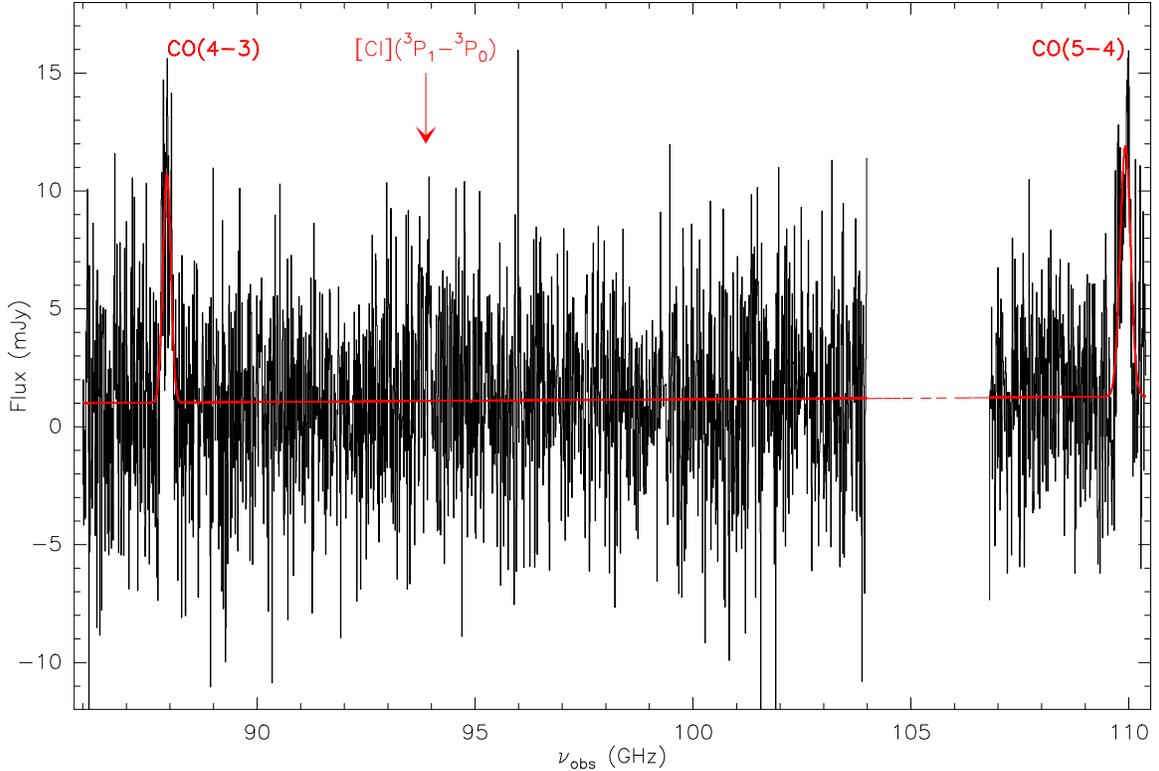,width=6in}}
\caption{The full frequency sweep done at the PdBI to search for CO emission lines in 
         ID\,141, showing the detections of CO($J=4\rightarrow 3$) and CO($J=5\rightarrow 4$). 
         The $4\rightarrow 3$ transition was detected first after covering 18~GHz in 5 frequency set-ups. 
         The $5\rightarrow 4$ emission line was detected thereafter in one setting, confirming the redshift 
         of the source at $z=4.243$. The line shows the continuum at a wavelength of 3~mm that is clearly detected 
         and is increasing from $\rm 1.2 \pm 0.1$ to $\rm 1.8 \pm 0.3 \, mJy$  from the lowest to the highest frequencies 
         -- as determined from the averaged line-free channels -- see Table~\ref{table:photometry}). 
         The position of the tentatively detected [C\,{\small I}]($^3P_1-^3P_0$) emission line is indicated.} 
\label{fig:freqsweep-PdB} 
\end{figure*}

We started the line search by tuning the central frequency of the band
to 98.6~GHz (96.8--100.4~GHz) with the aim to cover in a series of
observations the frequency range from 104.0 to 86.0~GHz that is
encompassing the redshift range ($3.5<z<4.5$) based on the probability
distribution of $z_{\rm phot}$. For this range in redshift, the
expected CO transition is the $4\rightarrow 3$.  For each setting, the
r.m.s.\ noise was $\rm 1\sigma \sim 4\, mJy$ in 10~MHz channels after
20~min on-source. After five frequency set-ups, covering 18~GHz from
104 to 86~GHz, a strong and broad emission line was detected at
87.9~GHz with a peak intensity of $\rm \sim 10 \, mJy$ and a zero
intensity width of $\rm \sim 1000 \, km s^{-1}$
(Fig.~\ref{fig:freqsweep-PdB} and Table~\ref{table:line-parameters}).

\begin{figure*} 
\centerline{\psfig{figure=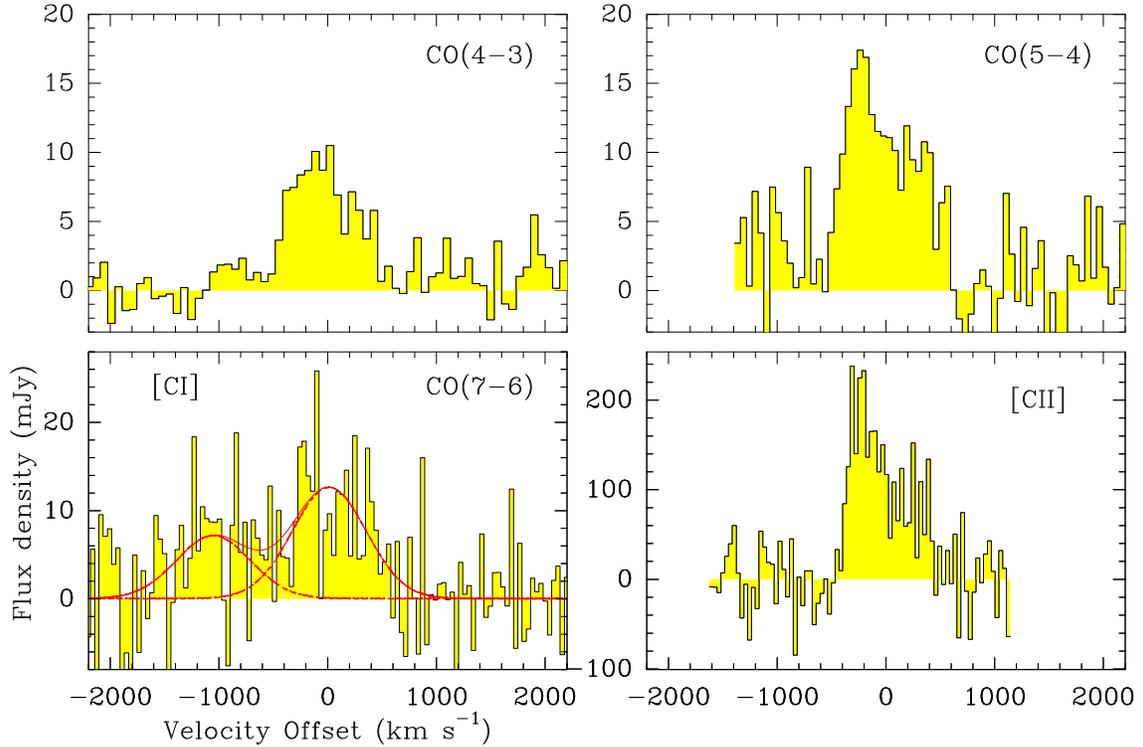,width=6in,angle=270}}
\caption{Spectra of the CO($J=4\rightarrow 3$) and CO($J=5\rightarrow 4$) (upper panels), the 
         CO($J=7\rightarrow 6$), [C\,{\small I}]($^3P_2-^3P_1$) and [C\,{\small II}]($^2P_{3/2}-^2P_{1/2}$) 
         emission lines (lower panels) towards ID\,141. The continuum was subtracted for all spectra.  
         The velocity scale corresponds to the frequencies ($\nu_{\rm obs}$) listed in 
         Table~\ref{table:line-parameters} for the CO and [C\,{\small II}] emission lines (i.e.\ for the redshift of
         $z=4.243$). Gaussian fits to the 
         CO($J=7\rightarrow 6$) and [C\,{\small I}] emission lines are shown as dashed curves whereas the 
         full line shows the composite fit.}
\label{fig:spectra} 
\end{figure*}

To identify the transition of the CO emission line detected at
87.9~GHz and search for the next higher CO transition, we tuned the
receivers to a central frequency of 108.5~GHz covering a redshift
range that includes the most probable case where the 87.9~GHz line is
the $4\rightarrow 3$ transition of CO. A strong line was immediately
apparent centered at 109.9~GHz, with a peak intensity of $\rm \sim 13
\, mJy$ and a broad line width comparable to that of the lower
frequency line. The two lines at 87.9 and 109.9~GHz are identified
with the $4\rightarrow 3$ and $5\rightarrow 4$ transitions of CO
(Figs.~\ref{fig:freqsweep-PdB} \& \ref{fig:spectra}).  These
identifications yield a redshift of $z=4.243 \pm 0.001$ for
ID\,141. This is the first 'blind' redshift determination using the
PdBI.

In addition to the CO emission lines, the 3-mm dust continuum emission
was detected at a level of $\rm 1.6 \pm 0.1 \, mJy$, averaged over the
entire line-free range covered at 3~mm, i.e.\ over $\rm \sim 19 \,
GHz$.  Table~\ref{table:photometry} lists the continuum flux densities
derived from this 3~mm spectral survey for three different
wavelengths (2.75, 3.00 and 3.29~mm) each averaged over $\rm \sim
4~GHz$.

Subsequent observations were performed on June 29 and September 11
2010, using the compact D configuration under moderate weather
conditions (PWV of 5--10~mm, $100< T_{\rm sys}{\rm /K} < 300$) to
search simultaneously for the emission lines of CO($7\rightarrow 6$)
and [C\,{\small I}]($^3P_2-^3P_1$) redshifted to 153.876 and
154.339~GHz. Both lines were detected with relatively good
signal-to-noise ratio together with the 2-mm dust continuum at a level
of $\rm 9.7 \pm 0.9 \, mJy$ (Fig.~\ref{fig:spectra}).  The detection
of the [C\,{\small I}]($^3P_2-^3P_1$) emission line prompted us to
check for the presence of the [C\,{\small I}]($^3P_1-^3P_0$)
transition that is redshifted to 93.87~GHz and is included in the
frequency sweep done from 104.0 to 86.0~GHz. Due to the low
signal-to-noise ratio of the sweep, the [C\,{\small I}]($^3P_1-^3P_0$)
line is detected with a poor signal-to-noise ratio and a line
intensity of $2.8 \pm 0.9 \, \rm Jy \, km \, s^{-1}$.

The parameters of the CO and [C\,{\small I}] lines observed towards
ID\,141 are listed in Table~\ref{table:line-parameters}.
Figure~\ref{fig:vel-int-maps} summarizes the results based on the PdBI
observations in a series of maps displaying both the continuum and
line emission.  The source, which is clearly not multiple, is barely
resolved in the PdBI data at the best available beam of
$3.6^{\prime\prime} \times 3.0^{\prime\prime}$ at 2~mm. In the 2-mm
continuum map, the integrated flux density of $\rm 9.7 \pm 0.9 \, mJy$
is higher than the emission peak of 5.5~mJy, and a {\it u-v} fit
yields a deconvolved size of $3.9\pm0.4 \times 2.4\pm 0.4 \, \rm
arcsec^2$ (or $(27 \times 16)/\mu_{\rm L} \, \rm kpc^2$) at an angle
of $\rm -25\pm10 \, deg$ (NE).
 
\subsection{SMA}

\begin{figure*}
\centerline{\psfig{figure=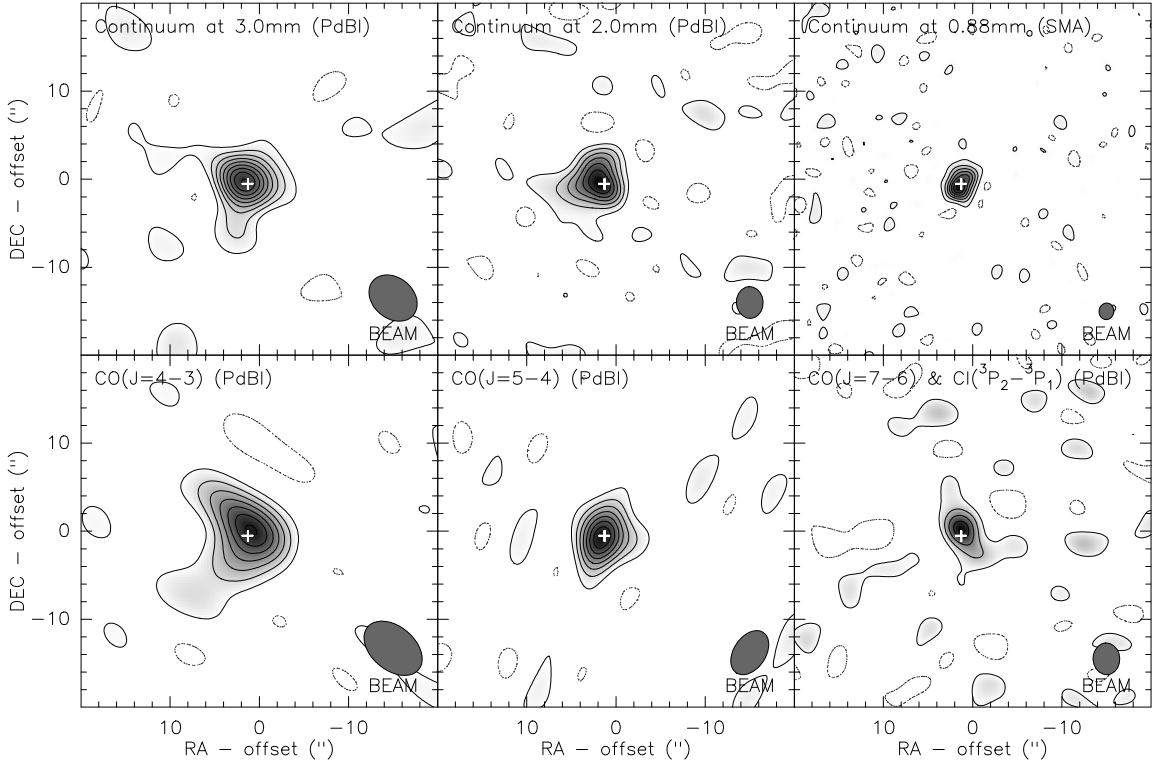,width=6in}}
\caption{{\it Upper panels}  Maps of the dust continuum emission at 3, 2 and 0.88~mm towards ID\,141. 
         {\it Lower panels}: Velocity-integrated maps of the three CO and [C\,{\small I}]($^3P_2-^3P_1$)
         emission lines towards ID\,141. All the data were obtained with the PdBI except the 0.88-mm
         continuum emission that was measured with the SMA. The contours are shown in steps of $2\sigma$, 
         starting at $\pm 2\sigma$ and the synthesized beams are shown in the right lower corner of each 
         panel. The $1\sigma$ values are: 0.09, 0.3 and 1.75~mJy/beam, respectively, for the 3-, 2- and 0.88-mm 
         continuum maps; 0.56, 0.75 and 1.21~$\rm Jy \, km \, s^{-1}/beam$, respectively, for the $4\rightarrow 3$, 
         $5\rightarrow 4$, and $7\rightarrow 6$ CO velocity-integrated maps. Offsets are given with respect 
         to the source position as determined by {\it Herschel} of RA 14:24:13.9, Dec.\ +02:23:04 (J2000). 
         The white cross indicates the position of the source as determined from the 0.88-mm map and 
         corresponds to RA 14:24:13.98, Dec.\ +02:23:03.45 (J2000).}
\label{fig:vel-int-maps} 
\end{figure*}

ID\,141 was observed in the continuum at 880~$\rm \mu m$ with the
Submillimeter Array (SMA) in its compact configuration with 8 antennas
on 2010 June 15 under moderate weather conditions. The observations on
source amounted to 2.4 hours.  The blazar 3C\,279 was utilized as a
bandpass calibrator, and Titan as the absolute flux calibrator.  The
resulting map is shown in Fig.~\ref{fig:vel-int-maps}.  The source,
which is well detected in the continuum, appears to be relatively
small with respect to the PdBI source size and is resolved at the
2$^{\prime\prime}$ resolution. The map r.m.s.\ noise is 1.75~mJy, with
a peak of $\rm \sim 50~mJy$.  The integrated flux density is $\rm 90
\pm 2 \, mJy$. To within the limits of the observations, ID\,141
appears to be elongated in the North-West direction, in agreement with
the 2-mm measurements of the PdBI.  The deconvolved size at 0.88~mm is
$2.4\pm0.1 \times 1.3\pm0.1 \, \rm arcsec^2$ (or $(16 \times
9)/\mu_{\rm L} \, \rm kpc^2$) at an angle of $\rm -32\pm4 \, deg$
(NE). The peak position of ID\,141, as derived from the 0.88-mm map,
corresponds to RA 14:24:13.98, Dec.\ +02:23:03.45 (J2000), which is
slightly offset by $1.3\arcsec$ from the position determined by
{\it Herschel} (see Fig.~\ref{fig:vel-int-maps}).  Clearly higher
angular resolution observations (both in the sub/millimeter and the
optical) are required to further explore the structure of ID\,141 and
its relation with the lensing galaxy.

\subsection{APEX}

Observations of the redshifted 1900.538~GHz C\,[{\small II}] line
emission in ID\,141 were made with the APEX 12-meter telescope in 2010
July in a series of two observing sessions, using the refurbished
dual-color FLASH+ MPIfR PI receiver. Its 345-GHz channel now operates
wideband IRAM 2SB mixers \cite{Maier2005}.  The observations took
place on July 7 (where the source was observed for 86 minutes on-time)
and July 25 (for 54 minutes on-time) during excellent weather
conditions with zenith precipitable water vapour of 0.7 and
0.4~mm. The spectra were taken in double-beam chopping mode, with a
wobbler throw of $50\arcsec$ at 1.5~Hz. Pointing was established on
nearby Mars and Saturn. At the frequency of 362.45~GHz, the beamsize
of APEX is $17.8\arcsec$.

The 4-GHz wide IF of the mixer was connected to a set of fast
Fourier-transform spectrometers. In the first observing session three
spectrometers from the CHAMP+FFT backend array, each 1.5~GHz wide,
were combined to process, with overlap, a total of 3.4~GHz. In the
second run -- during their commissioning -- two of the new
2.5-GHz-wide XFFT spectrometers were connected to cover the full 4~GHz
wide IF band (with 500~MHz of overlap), thereby providing $\rm 3300 \,
km \, s^{-1}$ instantaneous velocity coverage.

The [C\,{\small II}] emission line in ID\,141 was readily detected in
both observations. For data processing only linear offsets were
removed. The combined spectrum has an r.m.s.\ noise of 1~mK. The line
is detected with good signal-to-noise with a peak flux density of 4~mK
in $T_{\rm a}^*$ and a line shape that is comparable to that of the CO
emission lines (Fig.~\ref{fig:spectra}).  The total integrated
intensity of the [C\,{\small II}] line is $\rm 2.64\pm0.41\, K \, km
s^{-1}$ or $\rm 107\pm17 \, Jy \, km s^{-1}$
(Table~\ref{table:line-parameters}) -- at 362~GHz, the gain of the
APEX is 40.7 Jy/K in converting $T_{\rm a}^*$ to flux density
\cite{Gusten2006}. Because of the excellent weather conditions, the
absolute calibration uncertainties are expected to be of order
10--20\%.

The continuum at 850~$\rm \mu m$ was measured using the LABOCA
870-$\rm \mu m$ bolometer camera on 2010 September 20 for a total of
about 10~minutes under excellent weather conditions. Planets as well
as secondary calibrators were used for absolute flux calibration. The
source was detected with a flux density of $\rm 102.1\pm8.8 \, mJy$,
in good agreement with the SMA measurements
(Table~\ref{table:photometry}).  It is noted that the [C\,{\small II}]
contribution to the 60~GHz LABOCA passband amounts to only 2.1~mJy or
2\%, which is smaller than the uncertainties of the current 850-$\rm
\mu m$ continuum measurement.

\section{Discussion}
 
We have detected and/or mapped at high significance three high-$J$
emission lines of CO and the fine-structure emission lines of atomic
([C\,{\small I}]) and singly ionized ([C\,{\small II}]) carbon towards
ID\,141. Together with the detection and mapping of the far-infrared
dust continuum emission, these results are used to constrain the
properties of the molecular and atomic gas in this high-redshift
galaxy. The detection of the CO emission lines firmly establishes the
redshift of ID\,141 at $z=4.243\pm0.001$, a value within the redshift
range derived from the far-infrared/millimeter photometric
measurements.  This redshift corresponds to a luminosity distance $\rm
D_{\rm L} = 3.9 \times 10^4 \, Mpc$ and a scale of $\rm
6.9~kpc/^{\prime\prime}$.

\subsection{Far-infrared spectral energy distribution}

\begin{figure*}
\centerline{\psfig{figure=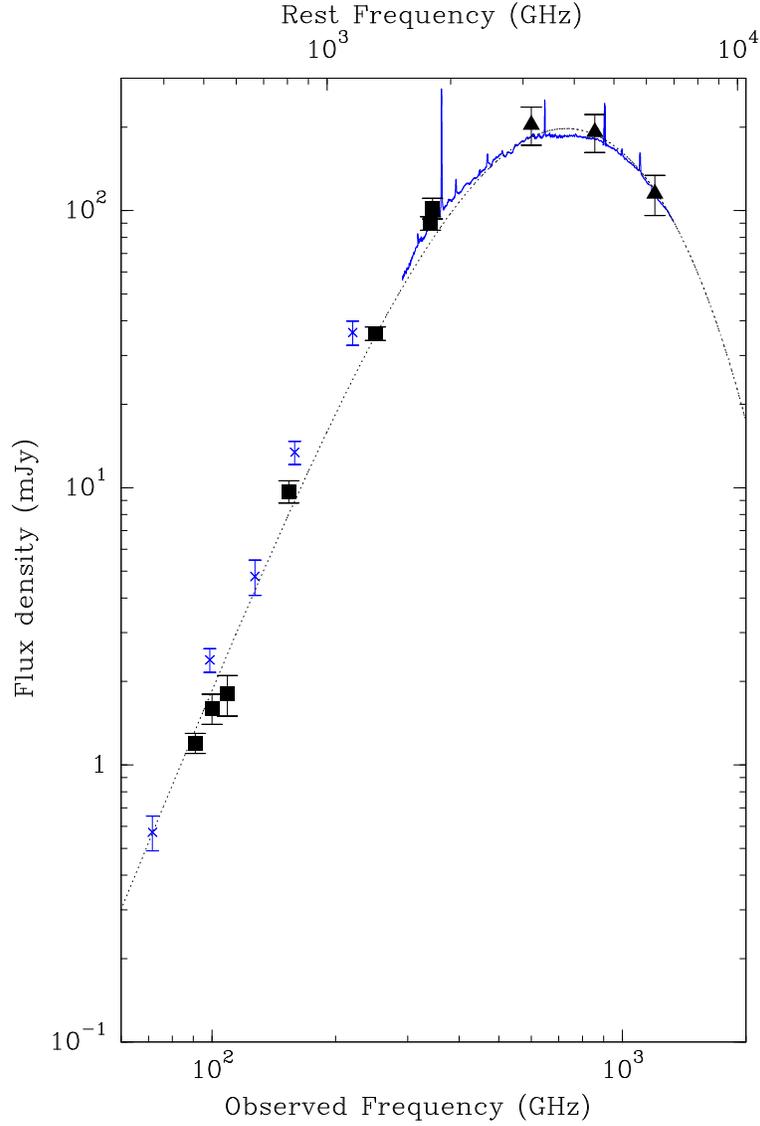,width=4in}}
\hspace{-0.5cm}
\caption{Spectral energy distribution of ID\,141 (all measurements from this paper: see Table~\ref{table:photometry}).
         The squares indicate the ground-based observations and the triangles the {\it Herschel} SPIRE data.  
         For comparison, the infrared SED of the starburst M~82 is displayed, redshifted to $z=4.24$ and normalized to the 
         flux density of ID\,141 at the observed wavelength of $\rm 500 \, \mu m$: the crosses show the photometric
         data taken from the literature and the continuous line is the ISO LWS spectrum of M~82 \cite{Colbert1999}. The dashed
         line displays the best fit to the photometric data of ID\,141 using a single component, optically thick
         dust model (as described in text).}  
\label{fig:sed}
\end{figure*}

Figure~\ref{fig:sed} displays the SED of ID\,141 -- based on the flux
densities listed in Table~\ref{table:photometry} -- where it is
compared with the SED of M82.  Both SEDs are comparable from the
rest-frame sub-millimeter to far-infrared wavelengths, both peaking at
a rest wavelength of $\rm \sim 90 \, \mu m$.  We fit the photometric
data points of ID\,141 with a single-temperature modified blackbody
spectrum (using the optically thin approximation) by performing a
$\chi^2$ fit on both the dust temperature ($T_{\rm dust}$) and the
dust emissivity index ($\beta$) as outlined in Beelen et
al.\ (2006). We derive $\beta = 1.7\pm 0.1$, $T_{\rm dust} = 38\pm 1
\rm \, K$ and a far-infrared luminosity of $L_{\rm FIR} \sim 8.5\pm0.3
\times 10^{13} \, \mu_{\rm L}^{-1} \, \rm L_\odot$, where $\mu_{\rm
  L}$ is the lensing magnification factor. To estimate the dust mass,
we adopt for the mass absorption coefficient $\kappa$ a value of $\rm
0.4 \, cm^2 \, g^{-1}$ at $\rm 1200 \, \mu m$ (e.g.\ Beelen et
al.\ 2006).  From the above dust temperature and far-infrared
luminosity, we derive $M_{\rm dust} \sim 8.9 \times 10^9 \, \mu_{\rm
  L}^{-1} \, \rm M_\odot$. Since the dust emission in luminous
infrared starbursts can be optically thick around $\rm 100 \, \mu m$,
we also modelled the SED of ID\,141 following the analysis described in
Wei\ss\ et al.\ (2007). A single component optically thick dust model
(adopting $\beta = 1.7$) fits the photometric data equally well as the
optical thin approximation, although with a higher dust temperature
($T_{\rm dust} \approx 58 \rm \, K$) and a lower dust mass ($M_{\rm
  dust} \sim 5.5 \times 10^9 \, \mu_{\rm L}^{-1} \, \rm M_\odot$). The
resulting fit is shown in Figure~\ref{fig:sed}. In this model, the
dust becomes opaque ($\tau = 1$) at a rest-wavelength of $\rm 150 \,
\mu m$ (corresponding to an observing frequency of 380~GHz) and the
apparent equivalent radius -- the true galaxy radius for an unlensed,
face-on, filled circular disk \cite{Weiss2007} -- is $r_0 \approx 720
\, \mu_{\rm L}^{-1} \rm \, pc$.

The global star-formation rate (SFR) can be estimated from the
far-infrared luminosity using the relation SFR $\approx \delta_{\rm
  MF} \, (L_{\rm FIR}/10^{10} \, \rm L_\odot) \, M_\odot \, yr^{-1}$
where $\delta_{\rm MF}$ is a function of the present mass composition
of the stellar population \cite{Omont2001}. For a range in
metallicities, initial mass functions and starburst ages of 10 to
100~Myr, $\delta_{\rm MF}$ ranges in between 0.8 and 2. Under the
assumption that the far-infrared luminosity of ID\,141 is dominated by
starburst activity, we estimate $0.7 < {\rm SFR} < 1.7 \times 10^4 \,
\mu_{\rm L}^{-1} \rm \, M_\odot \, yr^{-1}$, which is an upper limit
in case there is an AGN contribution to the far-infrared luminosity.
  
Using the FIRST radio survey \cite{Becker1995}, the 1.4-GHz flux
density of ID\,141 is measured to be $S_{\rm 1.4 \, GHz} = 570 \pm 160
\rm \, \mu Jy$, which shows that ID\,141 is not a radio-loud source.
For ID\,141, the far-infrared/radio parameter $q = {\rm
  log_{10}}[(F_{\rm IR}/3.75 \times 10^{12} {\rm \, W \,
    m^{-2}})/(S_{\rm 1.4 GHz}/{\rm W \, m^{-2} \, Hz^{-1}})]$
\cite{Condon1992} is estimated to be $q\sim 2.1\pm 0.1$, after
$K$-correcting and assuming a canonical radio slope for a star-forming
galaxy $\alpha=-0.75$, where $S_{\nu}\propto \nu^{\alpha}$. This value
is compatible with the median value $q \sim 2.4$ derived for
starbursts samples selected from BLAST or the {\it Herschel}
GOODS-North surveys (Ivison et al.\ 2010a, 2010b; Jarvis et
al.\ 2010). This suggests that star-formation activity is reponsible
for most of the far-infrared luminosity of ID\,141.

Typical amplification factors of $10-30$ have been estimated for
sources similar to ID\,141 with $\rm 500 \, \mu m$ flux density above
$\rm 100 \, mJy$ uncovered in the {\it H}-ATLAS survey
\cite{Negrello2010}.  We therefore adopt in this paper a magnification
range of $10< \mu_{\rm L} < 30$ for ID\,141. Under this assumption, the
far-infrared luminosity, dust mass and global star formation of ID\,141
are estimated to be in the range: 2.8 to $8.5 \times 10^{12} \, \rm
L_\odot$ for $L_{\rm FIR}$, 3 to $9 \times 10^8 \, \rm M_\odot$ for
$M_{\rm dust}$, and 200 to $1700 \, \rm M_\odot \, yr^{-1}$ for
$SFR$. These values are typical for the population of dusty, gas-rich
and luminous SMGs found at high redshift \cite{Greve2005,Tacconi2010}.
 
\subsection{The CO emission lines}

The CO lines of ID\,141 are broad ($\Delta V_{\rm FWHM} \sim 800 \, \rm
km \, s^{-1}$ - see Table~\ref{table:line-parameters}) and display a
clear, asymmetrical and double-peaked profile where the blue part of
the line is stronger than its red counterpart
(Fig.~\ref{fig:spectra}). A similar profile is seen in the [C\,{\small
    II}] emission line. The blue- and red-shifted peaks of the lines
are found from Gaussian fits at velocity offsets of $\pm 220 \, \rm km
\, s^{-1}$, respectively, each with a FWHM of $\sim \rm 400 \, km \,
s^{-1}$.  The broad line width of ID\,141 compares with the median FWHM
($780\pm320 \, \rm km \, s^{-1}$) of high-redshift SMGs and
double-peaked asymmetrical line shapes have also been reported for
other luminous submillimeter galaxies \cite{Greve2005,
  Swinbank2010}. Such double-peaked line profiles indicate orbital
motions under the influence of gravity either in the form of a disc or
a merger. However, the current angular resolution of the observations
presented in this paper does not allow to distinguish between the two
scenarios and attempts to find positional offsets between the blue-
and red-shifted peaks remained inconclusive. Higher resolution
observations in either CO or [C\,{\small II}] are required to further
examine the structure of the velocity gradient in ID\,141.

From the line parameters listed in Table~\ref{table:line-parameters}
and following the definition for line luminosity given by Eq.~(3) in
Solomon et al.\ (1997), the CO line luminosities are $L^{\prime} =
3.34\pm0.40$, $3.71\pm0.45$ and $0.94\pm0.20 \times 10^{11} \,
\mu_{\rm L}^{-1} \, \rm K \, km \, s^{-1} \, pc^2$ for the
$4\rightarrow 3$, $5\rightarrow 4$ and $7\rightarrow 6$ transitions,
respectively. This corresponds to
CO$(J=5\rightarrow 4)$/CO$(J=4\rightarrow 3)$ and
CO$(J=7\rightarrow 6)$/CO$(J=5\rightarrow 4)$ line brightness
temperature ratios of $r_{54}=1.1\pm 0.1$ and $r_{75}=0.24\pm 0.04$,
indicating that the higher CO transitions are not thermalized (see
also values of $L^{\prime}_{\rm CO}$ in
Table~\ref{table:line-parameters}).

\begin{figure*}
\centerline{\psfig{figure=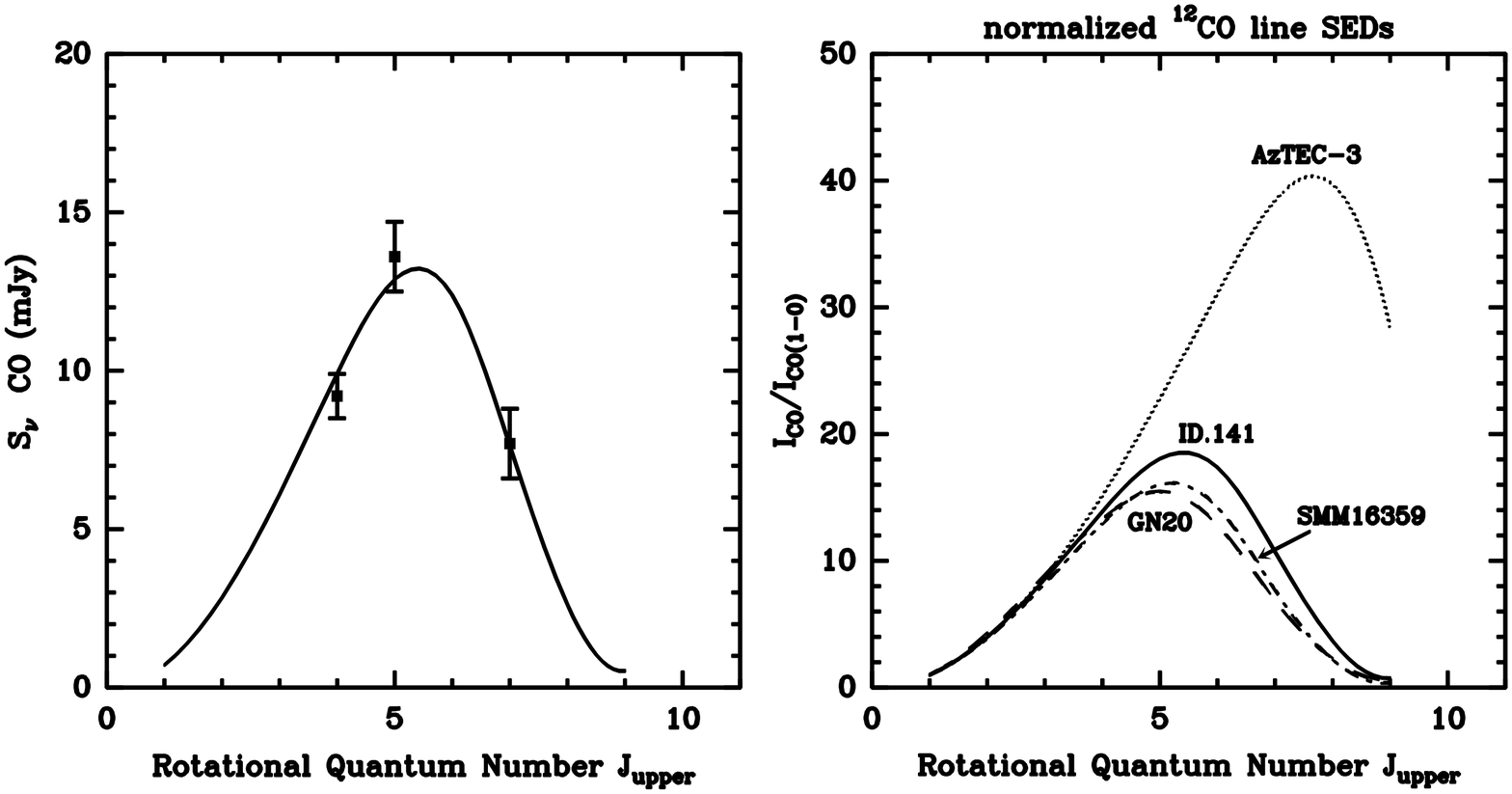,width=6in}}
\caption{{\it Left Panel} -- Observed CO fluxes vs.\ rotational quantum number (CO line SED) for ID\,141. The single-component 
         LVG model (described in the text) corresponds to $n({\rm H_2})=10^{3.9} \, \rm cm^{-3}$ and $T_{\rm kin} = 40 \, \rm K$. 
         {\it Right Panel} -- Comparison of the CO line SEDs of ID\,141 and selected high-redshift SMGs: GN20 ($z=4.05$) 
         \cite{Carilli2010}, SMM\,J16359 ($z=2.5$) \cite{Weiss2005b} and AzTEC-3 ($z=5.3$) \cite{Riechers2010}. The CO line 
         SEDs are normalized by their CO(1--0) flux density.}
\label{fig:LVG}
\end{figure*}

To describe the properties of the molecular gas in more detail, we
have analyzed the CO data using a spherical, single-component, large
velocity gradient (LVG) model \cite{Weiss2007}. As in Wei\ss\ et
al.\ (2007), we adopt a CO abundance per velocity gradient of $\rm
[CO]/(d{\it v}/d{\it r}) = 10^{-5} \, pc \, (km \, s^{-1})^{-1}$ that
best fits the high-$J$ lines and implies low enough opacities to
reproduce the fluxes of the two lower CO transitions.  The results of
the best-fit model, which are displayed in Fig.~\ref{fig:LVG},
correspond to an $\rm H_2$ density of $n_{\rm H_2} = 10^{3.9} \, \rm
cm^{-3}$, a gas temperature of $T_{\rm kin} = 40 \rm \, K$ and an
equivalent (magnified) radius of $r_0 = 680 \rm \, pc$. These values
are comparable to those derived for other high-redshift SMGs such as
GN20 at $z=4.04$ \cite{Carilli2010} and SMM\,J16359 at $z=2.5$
\cite{Weiss2005b}, as shown in Fig.~\ref{fig:LVG}.  Typical
degeneracies between the temperature and density in such LVG models
are discussed in Wei\ss\ et al.\ (2007).  Additional CO transitions,
in particular the low-$J$ transitions, should be observed to evaluate
more precisely the density and the temperature of the molecular gas
and check whether an excess of cold molecular gas could be present
\cite{Ivison2011}. Based on this model, the expected CO(1$-$0) line
luminosity is $L^{\prime}_{\rm CO(1-0)} = 4.3 \times 10^{11} \,
\mu_{\rm L}^{-1} \, \rm K \, km \, s^{-1} \, pc^2$ or $L_{\rm CO(1-0)}
= 2.1 \times 10^7 \, \mu_{\rm L} \, \rm L_\odot$.  Assuming a
conversion factor of $\alpha_{\rm CO} = 0.8 \, \rm M_\odot \, \rm (K
\, km \, s^{-1} \, pc^2)^{-1}$ from $L^{\prime}_{\rm CO(1-0)}$ to
$M_{\rm H_2}$, appropriate for starburst galaxies as derived from
observations of local ULIRGs \cite{Downes1998, Tacconi2008}, yields a
molecular gas mass $M_{\rm H_2} = 3.5 \times 10^{11} \, \mu_{\rm
  L}^{-1} \, \rm M_\odot$.  After correction for the amplification,
the above values are comparable to the median gas mass ($3 \times
10^{10} \, \rm M_\odot$) derived for luminous SMGs using the same
conversion factor $\alpha_{\rm CO}$ \cite{Greve2005} and the gas
masses of $(1-3) \times 10^{10} \rm \, M_\odot$ measured in the {\it
  H}-ATLAS high-redshift lensed SMGs, SDP.81 and SDP.130, discovered
by the {\it Herschel} space observatory \cite{Frayer2010}.

\subsection{The Carbon fine-structure atomic lines}

\subsubsection{The [C\,{\small II}] emission line}

From the total integrated intensity, the [C\,{\small II}] line
luminosity for ID\,141 is $L_{\rm C{\small II}} = 6.1\pm0.9 \times
10^{10} \, \mu_{\rm L}^{-1} \rm \, L_{\odot}$. Correcting for the
amplification factor, the value of $L_{\rm C{\small II}}$ for ID\,141
is therefore comparable to that of other high-redshift sources, such
as: SDSS~J1148+5251 ($z=6.42$) with $4.4\pm0.5 \times 10^{9} \rm \,
L_{\odot}$ \cite{Maiolino2005, Walter2009}, BR~0952$-$0115 ($z=4.43$)
with $4.6\pm0.7 \times 10^{9} \rm \, L_{\odot}$ \cite{Maiolino2009},
BRI~1335$-$0417 ($z=4.40$) with $16.4\pm2.6 \times 10^{9} \rm \,
L_{\odot}$ \cite{Wagg2010}, SMM~J2135$-$0102 (the `Cosmic Eyelash' at
$z=2.3$) with $5.5\pm1.3 \times 10^{9} \rm \, L_{\odot}$
\cite{Ivison2010c}, or the submillimeter galaxy, LESS~J033229.4
($z=4.76$) with $1.02\pm0.15 \times 10^{10} \rm \, L_{\odot}$
\cite{DeBreuck2011}.  The [C\,{\small II}] line luminosities for these
high-redshift infrared-luminous galaxies are, however, lower than the
most of the values derived for a sample of 14 galaxies at redshifts
$1<z<2$ that have bright [C{\small II}] emission lines yielding
luminosities in the range $8\times10^9 < L_{\rm C_{\small II}}/L_\odot
< 10^{11}$ \cite{Hailey-Dunsheath2010, Stacey2010}.
 
\begin{figure*}
\centerline{\psfig{figure=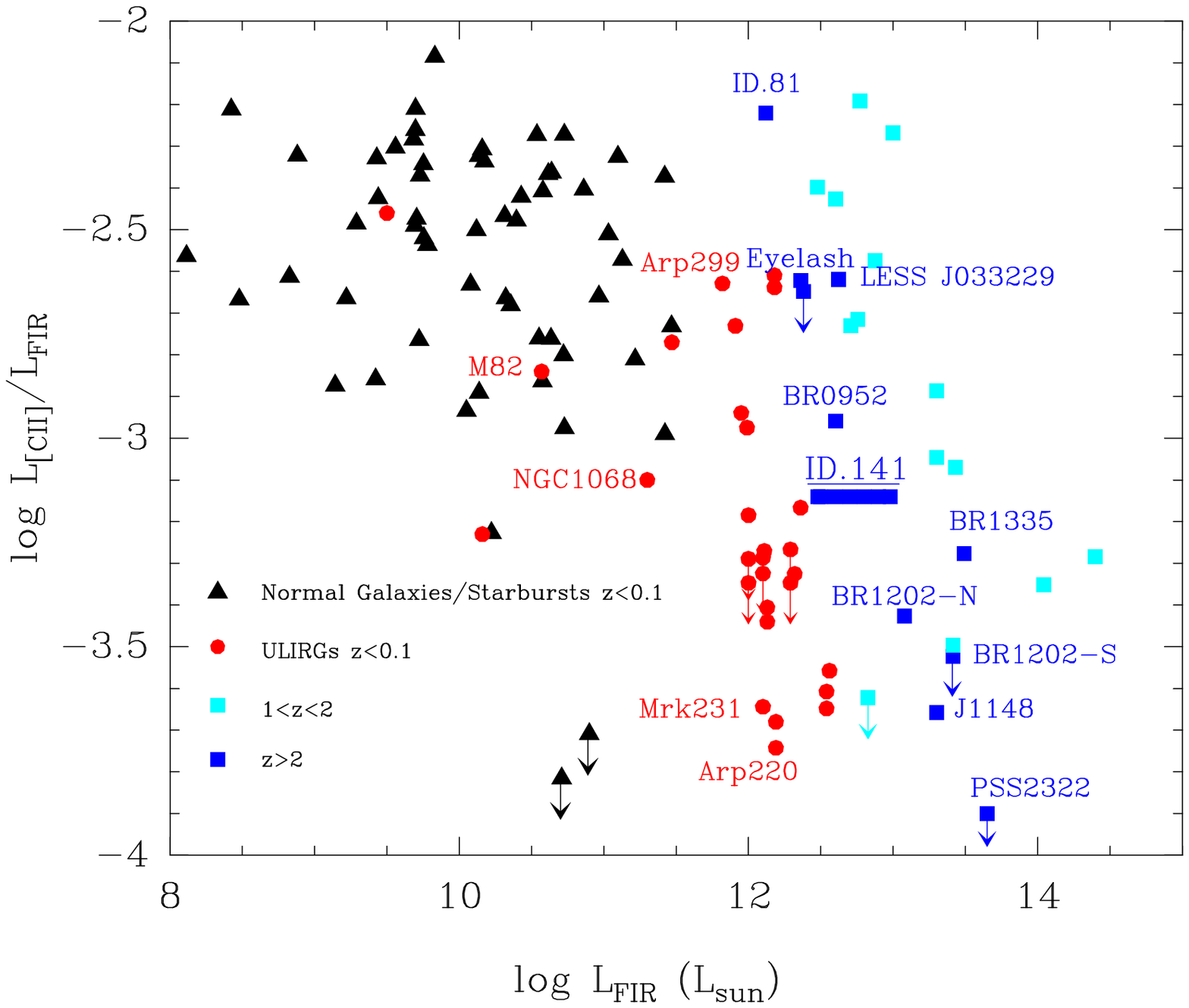,width=6in}}
\caption{The ratio $L_{\rm C{\small II}}/L_{\rm FIR}$ as a function of the far-infrared luminosity for normal galaxies 
         \cite{Malhotra2001,Colbert1999,Unger2000,Spinoglio2005,Carral1994} and ULIRGs \cite{Luhman1998,Luhman2003} in 
         the local universe, and starburst galaxies and AGN at $1<z<2$ \cite{Stacey2010} and $z>3$ -- see text 
         for references.  The source ID\,141 is labelled (and underlined) as well as other well-known local starbursts 
         (including M~82) and  high-redshift sources. The source ID.81 ($z=3.04$) is a recent measurement 
         based on SPIRE FTS observations (Valtchanov et al.\ 2011). For ID\,141, the bar shows the range in far-infrared
         luminosity after correcting for an amplification factor $10<\mu_{\rm L}<30$.}  
\label{fig:cii-fir}
\end{figure*}

For ID\,141, we derive a ratio $L_{\rm [C_{\small II}]} / L_{\rm FIR} =
(7.3\pm1.31) \times 10^{-4}$. This is higher than for most of the
high-redshift sources with similar infrared luminosity and redshift
but comparable to BRI~1335$-$0417 where the ratio is $(5.3\pm0.8)
\times 10^{-4}$ \cite{Wagg2010}. ID\,141 follows the trend reported for
high-redshift luminous galaxies in the relation of $L_{\rm [C_{\small
      II}]} / L_{\rm FIR}$ as a function of the far-infrared
luminosity $L_{\rm FIR}$, with a ratio of about one order of magnitude
lower than for normal galaxies but comparable to ULIRGs values
(Fig.~\ref{fig:cii-fir}).

Recently Graci\'a-Carpio et al.\ (2011) showed that the ratio between
the far-infrared luminosity and the molecular gas mass, $L_{\rm
  FIR}/M_{\rm H_2}$, appears to be a better choice from a physical
point of view to investigate the relative brightness of the
far-infrared fine-structure lines in galaxies. This ratio, which
removes the uncertainty of the amplification factor, also appears to
better reflect the properties of the gas than $L_{\rm FIR}$
alone. Graci\'a-Carpio et al.\ (2011) found that there is a threshold
at $L_{\rm FIR}/M_{\rm H_2} \sim 80 \, \rm L_\odot \, M_\odot^{-1}$
above which galaxies tend to have weaker [C\,{\small II}] emission
with respect to their far-infrared luminosity. This threshold is
similar to the one separating the star-formation relations between
non- or weakly-interacting star-forming galaxies and luminous mergers
\cite{Genzel2010, Daddi2010}. With a ratio $L_{\rm FIR}/M_{\rm H_2}
\sim 230 \, \rm L_\odot \, M_\odot^{-1}$, ID\,141 follows the trend in
the $L_{\rm [C_{\small II}]} / L_{\rm FIR}$ vs. $L_{\rm FIR}/M_{\rm
  H_2}$ plane reported in Graci\'a-Carpio et al.\ (2011), indicating
that its ionization parameter must be high ($\sim 10^{-2}$) and that
ID\,141 could be a luminous merger.

Another sensitive probe of the physical conditions within the
interstellar medium is the ratio $L_{\rm C{\small II}}/L_{\rm
  CO(1-0)}$.  For ID\,141, we find from the measured [C\,{\small II}]
and the estimated CO(1--0) luminosities, $L_{\rm C{\small II}}/L_{\rm
  CO(1-0)} \sim 2900$. Using the Photon-Dominated Region (PDR) model
and diagnostic diagram comparing the ratios $L_{\rm C_{\small
    II}}/L_{\rm FIR}$ versus $L_{\rm CO(1\rightarrow0)}/L_{\rm FIR}$
\cite{Kaufman1999} indicates that, for ID\,141, the gas is dense with
$n \sim 10^4 \, \rm cm^{-3}$ and that the far-UV radiation field
illuminating the PDR, $G_0$, is a few $10^3$ times more intense than
that in our Galaxy. Both these values point to physical conditions
comparable to those found in the central regions of local ULIRGs
\cite{Stacey2010}.

\subsubsection{The [C{\small I}] emission line}

The [C\,{\small I}]($^3P_2-^3P_1$) line flux implies $L_{\rm C{\small
    I}} = 0.8\pm0.3 \times 10^{9} \, \mu_{\rm L}^{-1} \rm \,
L_{\odot}$ (Table~\ref{table:line-parameters}).  The [C\,{\small I}]
luminosity in ID\,141, once corrected for the amplification, is
therefore lower than the [C\,{\small I}] luminosities derived from the
$(^3P_2-^3P_1)$ transition for other high-redshift sources such as the
Cloverleaf \cite{Weiss2003, Weiss2005a} or PSS~2322+1944
\cite{Pety2004} or other high-redshift galaxies or quasars
\cite{Walter2011}. Assuming an excitation temperature ($T_{\rm ex}$)
equal to the temperature of the dust and gas, i.e.\ $T_{\rm ex} \sim
40 \, \rm K$, the mass of neutral carbon in ID\,141 amounts to $M_{\rm
  C{\small I}} =6.5\pm1.1 \times 10^7 \, \mu_{\rm L}^{-1} \, \rm
M_\odot$, a result that is only weakly dependent on $T_{\rm ex}$
\cite{Pety2004}. Compared to the mass of molecular gas, the derived
mass of [C\,{\small I}] implies a neutral carbon abundance relative to
$\rm H_2$ of $\rm [C{\small I}]/[H_2] \approx 1.8 \times 10^{-5}$.
The similarity of this abundance with the value of $2.2 \times
10^{-5}$ found for Galactic dense molecular clouds \cite{Frerking1989}
is indicative of a substantial enrichment in ID\,141. In addition, the
$L_{\rm C{\small I}(1-0)}/L_{\rm FIR}$ luminosity ratio is also a
sensitive measurement of $G_0$ \cite{Kaufman1999, Gerin2000}. For
ID\,141, $L_{\rm C{\small I}(1-0)}/L_{\rm FIR} \approx 5 \times
10^{-6}$ close to the luminosity ratio found in other high-redshift
sources \cite{Walter2011} and indicating a high UV field strengh of
$G_0 \sim 3 \times 10^3$, compatible with the analysis based on the
[C\,{\small II}] emission line.

The measurements available for ID\,141 allow an estimate of the
relative cooling of the carbon and CO emission lines.  The upper
transition of [C\,{\small I}] is stronger than the lower transition by
a factor of $\approx 2$ (Table~\ref{table:line-parameters}), as in the
case of the Cloverleaf \cite{Weiss2005a}. The total luminosity of
ID\,141 in [C\,{\small I}] is therefore $\approx 1.2 \times 10^9 \,
\mu_{\rm L}^{-1} \rm \, L_\odot$. Based on the LVG model described
above, the total CO line luminosity in ID\,141 is estimated to be
$L_{\rm CO} \approx 8.4 \times 10^9 \, \mu_{\rm L}^{-1} \rm \,
L_\odot$. The CO/[C\,{\small I}] luminosity ratio is thus $\approx
7.0$ and the cooling in the CO and [C\,{\small I}] lines represents
$10^{-4}$ of the far-infrared continuum, a factor of 7 lower than the
main cooling line of the gas, i.e the [C\,{\small II}] transition.

\section{Conclusions}

We have used four ground-based (sub)millimeter facilities -- the IRAM
30-meter and PdBI, the SMA and APEX -- to follow up the exceptional
source ID\,141 that is the strongest 500-$\rm \mu m$ riser yet
discovered in the {\it Herschel} cosmological surveys. The detection
of three bright CO emission lines using the PdBI confirms that the
source is at high redshift ($z=4.243$). The strength of the continuum
and the intensities of the molecular and atomic emission lines of
ID\,141 suggest the source is gravitationally lensed, in line with what
was already found for other sub-millimeter bright, high-redshift,
galaxies detected with the {\it Herschel} space observatory
\cite{Negrello2007, Negrello2010}. The magnification factor is still
unknown but is likely to be in the range $10 < \mu_{\rm L} <30$ as
found for other gravitationally lensed SMGs uncovered in the {\it
  H}-ATLAS survey.  The properties of the molecular and atomic gas in
ID\,141 indicate properties of a luminous ($L_{\rm FIR} \sim 3 - 8
\times 10^{12} \, L_\odot$), dense ($n \approx 10^4 \, \rm cm^{-3}$)
and warm ($T_{\rm kin} \approx 40 \rm \, K$) starburst galaxy that are
comparable to those derived for local starburst galaxies or
high-redshift submillimeter galaxies. The asymmetric double profiles
of the carbon and CO emission lines are indicative of orbital motions
in a disc or a merger. The source is barely resolved in the present
data indicating a physical size that is smaller than $14/\mu_{\rm
  L}$~kpc, corresponding to an upper limit of $2^{\prime\prime}$.

The measurements of the carbon lines ([C\,{\small I}] and [C\,{\small
    II}]) in ID\,141 enable a first study of the gas associated with
the PDR in ID\,141 and analysis of the relative cooling of the carbon
and CO emission in this galaxy, that is dominated by that of $\rm
C^+$. The [C\,{\small II}]-to-FIR luminosity ratio, $(7.3 \pm 1.3)
\times 10^{-4}$, is larger than the one found for other galaxies of
comparable luminosity and redshift, and consistent with recent
findings that the scatter in the $L_{\rm C{\small II}}/L_{\rm FIR}$
versus $L_{\rm FIR}$ diagram is indeed large. With a ratio $L_{\rm
  FIR}/M_{\rm H_2} \sim 230 \, \rm L_\odot \, M_\odot^{-1}$, ID\,141
follows the trend reported by Graci\'a-Carpio et al.\ (2011) in the
$L_{\rm C{\small II}}/L_{\rm FIR}$ versus $L_{\rm FIR}/M_{\rm
  M_{H_2}}$ plane, suggesting both a high ionization parameter and the
fact that ID\,141 could be a merger.

The present observations underline the importance of observations done
at frequencies in between 85 and 360~GHz to study the properties of
galaxies at redshift $z>4$. In particular, the lower frequencies (in
the 3 and 2~mm windows) are essential to probe the peak of the CO
emission that, for such starburst galaxies, lies around $J_{\rm
  upper}$ 4 to 5, as well as the two emission lines of [C\,{\small I}]
or molecules other than CO such as water. Follow-up observations of
very-high-redshift, obscured galaxies found in future deep surveys,
especially the less luminous objects, will need sensitive
interferometers operating at frequencies below 350~GHz such as ALMA,
the EVLA or the future upgrade of the Plateau de Bure interferometer
(NOEMA) in order to fully explore the properties of the molecular and
atomic gas in starburst galaxies that were active when the universe
was about 1~Gyr old.

\acknowledgments

The results described in this paper are based on observations obtained
with {\it Herschel}, an ESA space observatory with science instruments
provided by European-led Principal Investigator consortia and with
important participation from NASA. US participants in {\it H}-ATLAS
acknowledge support from NASA through a contract from JPL. The
ground-based follow-up observations were obtained at the following
facilities. The 30-meter telescope and the PdBI of IRAM that is funded
by the Centre National de la Recherche Scientifique (France), the
Max-Planck Gesellschaft (Germany) and the Instituto Geografico
Nacional (Spain). APEX is a collaboration between the
Max-Planck-Institut f\"ur Radioastronomie, the European Southern
Observatory, and the Onsala Space Observatory. The Submillimeter Array
(SMA) is a joint project between the Smithsonian Astrophysical
Observatory and the Academia Sinica Institute of Astronomy and
Astrophysics and is funded by the Smithsonian Institution and the
Academia Sinica. Part of this study was supported by financial
contribution from the agreement ASI-INAF I/009/10/0 and by CONACyT
grant 39953-F.

{\it Facilities:}  \facility{{\it Herschel} (SPIRE)}, \facility{IRAM 30-meter}, \facility{IRAM PdBI}, \facility{SMA}, \facility{APEX}.

\bibliographystyle{apj} 

{}

\newpage



\begin{table}
\caption{Photometric Measurements of ID\,141}
\label{table:photometry}
\vspace{0.8cm}
\centering{
\begin{tabular}{ccl}
\hline \noalign{\smallskip}
\hline \noalign{\smallskip}
 Wavelength  & Flux Density  & Facility \\
 $\rm [\mu m]$     & [mJy]        &    \\
\noalign{\smallskip} \hline \noalign{\smallskip}                   
      250 & 115$\pm$19 & SPIRE$^{1}$ \\
      350 & 192$\pm$30 & SPIRE$^{1}$ \\
      500 & 204$\pm$32 & SPIRE$^{1}$ \\
      870 & 102$\pm$8.8 & LABOCA$^{2}$ \\
      880 &  90$\pm$5   & SMA \\
     1200 &  36$\pm$2 & MAMBO-2$^{3}$ \\
     1950 &  9.7$\pm$0.9 & PdBI \\
     2750 &  1.8$\pm$0.3 & PdBI \\
     3000 &  1.6$\pm$0.2 & PdBI \\
     3290 &  1.2$\pm$0.1 & PdBI \\
\noalign{\smallskip} \hline
\end{tabular}\\ 
\vspace{0.4cm}
Note -- $^{1}$ {\it Herschel}; $^{2}$ APEX; $^{3}$ 30-meter. 
}
\end{table}


\begin{table}
\caption{Parameters of the CO, [C\,{\small I}] and [C\,{\small II}] emission lines observed towards ID\,141}
\label{table:line-parameters}
\vspace{0.8cm}
\centering{
\begin{tabular}{lcccccl}
\hline \noalign{\smallskip}
\hline \noalign{\smallskip}
 Line    & $\nu_{\rm obs}$ & Peak Int. & $\rm \Delta V_{\rm FWHM}$  &     $I$         &         $L$  &  $L'$  \\
& [GHz] & [mJy] & [$\rm km \, s^{-1}$] & [Jy km~s$^{-1}$] & [$10^9 \, L_{\odot}$] & [$\rm 10^{10} \, K \, km \, s^{-1} \, pc^2$]\\
\noalign{\smallskip} \hline \noalign{\smallskip}                   
CO(4--3) &  87.929    &   9.2$\pm$0.7   &  760$\pm$70   &   7.5$\pm$0.9  &     1.0$\pm$0.1 & 33.4$\pm$4.0  \\   
CO(5--4) & 109.911    &  13.6$\pm$1.1   &  890$\pm$80   &  13.0$\pm$1.6  &     2.3$\pm$0.3 & 37.1$\pm$4.5  \\
CO(7--6) & 153.876    &   7.7$\pm$1.2    & 790$\pm$130  &   6.5$\pm$1.4  &     1.6$\pm$0.3 & 9.4$\pm$2.0  \\
\noalign{\smallskip}
$\rm [C\,{\small I}](^3P_1-^3P_0)$ &   93.870  &  3.0$\pm$1.0  & 790$^\dagger$  &   2.8$\pm$0.9 & 0.42$\pm$0.15 & 10.9$\pm$3.5  \\
$\rm [C\,{\small I}](^3P_2-^3P_1)$ &  154.339  &  4.0$\pm$1.0  & 790$\pm$130  &   3.4$\pm$1.1 & 0.8$\pm$0.3 & 4.9$\pm$0.9  \\
\noalign{\smallskip}
$\rm [C\,{\small II}](^2P_{3/2}-^2P_{1/2})$ & 362.45 &  147$\pm$20  & 690$\pm$80 &  107$\pm$17  & 61.6$\pm$9.8 & 28.1$\pm$4.46  \\ 
\noalign{\smallskip} \hline
\end{tabular}\\ 
\vspace{0.4cm}
Note -- The line luminosities are not corrected for lensing magnification. The amplification factor 
is supposed to be in the range $10 <\mu_{\rm L} < 30$ - see discussion in Sect.~3. $^\dagger$ Fixed linewidth. 
}
\end{table}

\newpage

\end{document}